\documentclass[fleqn,twoside]{article}
\usepackage{espcrc2}

\usepackage{epsfig}
\usepackage[figuresright]{rotating}

\newcommand{\AmS}{{\protect\the\textfont2
  A\kern-.1667em\lower.5ex\hbox{M}\kern-.125emS}}

\def\frac#1#2{ {{#1} \over {#2} }}

\def\beq{\begin{equation}}
\def\eeq{\end{equation}}

\def\non{\nonumber}
\def\beqn{\begin{eqnarray}}
\def\eeqn{\end{eqnarray}}

\def\L{\Lambda}

\def\as{\alpha_{\sf s}}
\def\a0{\alpha_{\sf 0}}
\def\aV{\alpha_{\sf V}}

\def\dm{\delta m}

\def\MSbar{\overline{\rm MS}}

\title{The $n_f=2$ residual mass in lattice HQET 
to $\alpha_s^3$ order\thanks{Talk presented by F. Di Renzo}
\vskip-3.6cm\hfill\small HU-EP-04/56; SFB/CPP-04-51; UPRF-2004-17\vskip3.3cm}

\author{F. Di Renzo\address{Dipartimento di Fisica, Universit\`a di 
														Parma and INFN, Gruppo Collegato di Parma, Italy} 
        and L. Scorzato\address{Instit\"ut f\"ur Physik, Humboldt Universit\"at, Berlin,                                    Germany}}
       
\begin{document}

\begin{abstract}
We compute the so called residual mass in Lattice Heavy Quark Effective Theory to $\alpha_s^3$ order in the $n_f=2$ (unquenched) case. The control of this additive mass renormalization is crucial for the determination of the heavy quark mass from lattice simulations. 
We discuss the impact on an unquenched determination of the b-quark mass. \vspace{1pc}
\end{abstract}

\maketitle

\section{Introduction}

The computation of quark masses is one of the most successfull application of Lattice QCD. Within this framework an interesting role is played by the determination of 
the $b$--quark mass. Despite the fact that one can not accomodate the $b$--quark on a lattice, its mass is computed to a very good accuracy. As it is known, the key of the success is the 
use of some form of effective theory. Many approaches have been used, whose results are consistent within errors in the quenched approximation \cite{NRQCD,Rm1HQET,AlphHQET,Rm2,Gunnar}. A popular determination comes from Heavy Quark Effective Theory (HQET). It shares with other approaches the need of dealing with the mass counterterm we will be concerned with. Within HQET, the most direct relation between the mass of a physical hadron ($M_B$) and the mass of the heavy quark is 
\begin{equation}
	M_B =  m_b + {\cal E} + {\cal O}(1/m_b),
\end{equation}
in which $m_b$ is the HQET expansion mass 
parameter and ${\cal E}$ is the (linearly divergent) binding energy. 
However, $m_b$ is not yet properly defined at this stage. A procedure for actually making use of the previous formula is implemented in \cite{Rm1HQET}. First of all, one matches the QCD propagator to its lattice HQET counterpart to obtain a relation involving the pole mass 
\begin{equation}
	m_b^{pole} = M_B - {\cal E} + \dm + {\cal O}(1/m_b).
\end{equation}
This relation contains the quantity whose perturbative computation we are reporting on, \emph{i.e.} the so called residual mass $\dm$, a linearly divergent 
additive mass counterterm which is peculiar to a hard cut-off regularization scheme like the lattice. The pole mass can in turn be related to the $\MSbar$ mass $\overline{m_b}=m_b^{\MSbar}$
\begin{equation}
	\overline{m_b}(\overline{m_b}) = 
m_b^{pole} \, 
\left [ 1 + \sum_{n=0}^{\infty} (\frac{\as(\overline{m_b})}{\pi})^{n+1}
D_n \right ] \;,
\end{equation}
the $D_n$ coefficients with $n \leq 2$ being known from \cite{CSR99,MR99}. If we are now able to compute ($\alpha_0$ is the lattice coupling) 
\begin{equation}
	\dm  =  \sum_{n \geq 0} \, \overline{X_n} \, \a0^{n+1}
\end{equation}
one can put everything together to get
\begin{eqnarray}
	\overline{m_b}(\overline{m_b}) & = & 
\left [ M_B - {\cal E} + \sum_{n=0}^{\infty} (\as)^{n+1}
\frac{X_n}{a}\right ] \times \nonumber \\  
& & \left [ 1 + \sum_{n=0}^{\infty} (\frac{\as}{\pi})^{n+1}
D_n \right ] \;.
\end{eqnarray}
In this relation $\as$ is $\as(\overline{m_b})$ (and the $\overline{X_n}$ get thus translated into $X_n$). ${\cal E}$ can be computed as the decay constant of the correlation function of two axial currents. Knowing $\dm$ is crucial since it cancels both the divergence of ${\cal E}$ and the renormalon ambiguities which come from the matching between $m_b^{pole}$ and $m_b^{\MSbar}$ \cite{MS95}. Since both cancelations take place in Perturbation Theory, one has to carefully assess to which accuracy one can control the procedure. While $X_0$ has been known for a long time, $X_1$ was computed in \cite{MS98}, while $X_2^{(n_f=0)}$ was computed in \cite{PR01} and then also in \cite{USApt}. The knowledge of $X_2^{(n_f=0)}$ greatly improved the accuracy of the quenched determination of the $b$--quark mass. 

\section{The computation}

$X_2^{(n_f=2)}$ was computed using the same strategy used for $X_2^{(n_f=0)}$ \cite{PR01}. We computed the Wilson loops $W(R,T)$ for all the values of $R$ and $T$ up to $16$. This was done on a $32^4$ lattice using Numerical Stochastic Perturbation Theory. The computation was done for Wilson gauge and Wilson quarks action and the sea quarks masses were put to zero (\emph{i.e.} the appropriate counterterms for the perturbative critical mass were plugged in). From Wilson loops and Creutz's ratios one can compute
\begin{equation}
	V_T(R) \equiv \log \left( \frac{W(R,T-1)}{W(R,T)} \right)
\end{equation}
which in turn can yield the static potential via
\begin{equation}
	V(R) \equiv \lim_{T\rightarrow\infty} V_T(R).
\end{equation}
The static potential is just the sum of the Coulomb potential (which can be described in terms of a potential coupling in whose definition the logarithmic divergencies\footnote{In Wilson loops also the so called corner logarithmic divergencies are present, but they cancel in the Creutz's ratios.} are absorbed) and the (linearly divergent) mass counterterm we are interested in \cite{DV80}
\begin{equation} \label{matching}
	V(R) = 2 \, \dm \, - \, C_F \frac{\aV(R)}{R}.
\end{equation}
This means that the perturbative computation of the static potential can be read as the computation of the matching between the lattice and the potential couplings, \emph{i.e.} in the previous formula 
\begin{equation}
	\aV(R) =  \a0 + c_1(R) \, \a0^2 + c_2(R) \, \a0^3 + \, \ldots,
\end{equation}
the coefficients $c_1(R)$ and $c_2(R)$ being dictated by $\Lambda$--parameters and $\beta$--functions coefficients 
\[
	c_1(R) =  2 b_0 \log R + 2 b_0 \log \frac{\L_V}{\L_0}  
\]
and
\begin{eqnarray}
c_2(R) & = & {c_1(R)}^2 + 2 b_1 \log R + 2 b_1 \log \frac{\L_V}{\L_0} + \non \\
& & + \frac{b_2^{(V)}-b_2^{(0)}}{b_0}. \non
\end{eqnarray}
Since the matchings between both the potential and the $\MSbar$ couplings \cite{York} and the $\MSbar$ and the lattice couplings \cite{LMS} are known, the only unknown quantity in Eq.~(\ref{matching}) is $\dm$. Our procedure was to fit the coefficients $X_n^{(n_f=2)}$ by matching our perturbative computation of the static potential to Eq.~(\ref{matching}). An example of the fitting procedure can be seen in Fig.~1. The intervals in which fits were performed were such that: $R\geq3$, $T\geq12$ and $T > 2.5 \overline{R}$ ($\overline{R}$ is the mean value of $R$ in the fitting interval; the fitting intervals themselves were from $3$ up to $7$ points long). We then choose in terms of $\chi^2$: the errors we quote refer to the interval embraced by letting $\chi^2$ vary within a given interval. On top of that we could also inspect that the impact of lattice artifacts was under control (an handle is the value of known parameters entering the matching relations).

\section{Results}
We got $X_0=2.118(2)$, $X_1^{(n_f=2)}=10.56(4)$ (to be compared with the analytical values $X_0=2.118...$ and \cite{MS98} $X_1^{(n_f=2)}=10.588...$) and finally
\begin{equation}
	X_2^{(n_f=2)} = 76.7(6).
\end{equation}
Using the last result the authors of \cite{Rm1HQET} were able to repeat the analysis of their $n_f=2$ unquenched data. The result for the $b$--quark mass is \cite{ViceRak}
\[
\overline{m_b}(\overline{m_b})^{(\mbox{unq})} = (4.21 \pm 0.03 \pm 0.05 \pm 0.04) \;\; \mbox{GeV}.
\]
At least three considerations are in order. First of all, by taking into account the value of $X_2^{(n_f=2)}$ the central value moved from $4.26$ to $4.21 \mbox{GeV}$ and (what is more important) the last error, which is the one taking into account the indeterminations in the perturbative 
matching, got roughly halved. A second observation is that the new analysis yields different values for 
the quenched and unquenched results. Still, they are compatible 
within errors and in the end further investigation is needed. The third point is the assessment of 
sistematic errors involved in the procedure. In this approach one sticks to finite values of the lattice spacing. The finite lattice spacing dependence is not dramatic and it gets decreased by including the new term in the matching. On the other hand, the control on renormalon ambiguities seems firm. \\
After the conference the analysis was refined and completed. The final version of the computation can be found in \cite{PR04}.

\begin{figure}[t]
\vspace{9pt}
\begin{center}
\mbox{\epsfig{figure=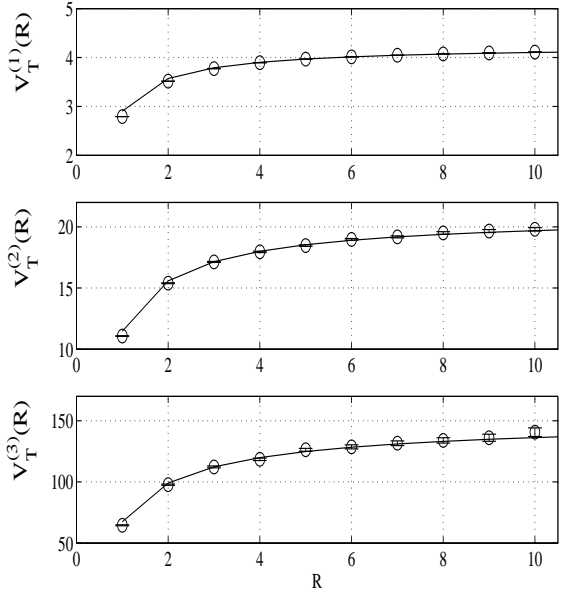,width=7.5cm,height=8cm}}
\caption{The result of a fitting procedure as described in the text. Solid lines are the content of Eq.~(\ref{matching}) once $\dm$ has been fitted, while circles are the computed $V_T^{(i)}(R)$, \emph{i.e.} orders $\alpha_0^i$ of $V_T(R)$. In these figures $T=15$.}
\end{center}
\end{figure}

\section*{Acknowledgments}

We thanks V. Gimenez which on behalf of the authors of \cite{Rm1HQET} shared with us the results of their new analysis of the $b$--quark mass, with our new result taken into account.

\end{document}